\documentclass[aps,pra,twocolumn,showpacs,groupedaddress]{revtex4-1}

\usepackage{amsmath}
\usepackage{epsfig}

\begin{document}  
\title {\bf Full characterization of modular values for two-dimensional systems}
\author{ Le Bin Ho}
\thanks{Electronic address: binho@qi.mp.es.oaska-u.ac.jp}
\affiliation{Department of Materials Engineering Science, Graduate School of Engineering Science, Osaka University, Toyonaka, Osaka 560-8531, Japan}

\author{ Nobuyuki Imoto}
\affiliation{Department of Materials Engineering Science, Graduate School of Engineering Science, Osaka University, Toyonaka, Osaka 560-8531, Japan}

\date{\today}

\begin{abstract}
Vaidman pointed out the importance of modular values, and related the modular value of a Pauli spin operator to its weak value for specific coupling strengths [Phys. Rev. Lett. \textbf{105}, 230401 (2010)]. It would be useful if this relationship is generalized since a modular value, which assumes a finite strength of the measurement interaction, is sometimes more practical than a weak value, which assumes an infinitesimally small interaction. In this paper, 
we give a general expression that relates the weak value and the modular value of an arbitrary observable in the 2-dimensional Hilbert space for an arbitrary coupling strength. Using this expression, we show the ^^ ^^ failure of sum rule" for modular values, which has a resemblance to the ^^ ^^ failure of product rule" for weak values. We give examples of ^^ ^^ failure of sum rule" for some interesting cases, i.e., paradoxes based on nonlocality, which include EPR paradox, Hardy's paradox, and Cheshire cat experiment. 
\end{abstract}
\pacs{03.65.Ta, 03.65.Ud, 02.10.Yn, 02.60.Ed}
\maketitle

{\section{INTRODUCTION}}
Weak value is a groundbreaking concept discovered by Aharonov, Albert, and Vaidman \cite{AAV}.
A weak value of an observable $\hat{A}$ is defined to be the expectation value of $\hat{A}$ by a weak measurement performed between the pre-selection of an initial state and the post-selection of a final state. Counter to usual expectation values, weak values can far outside the range of eigenvalues of the observable $\hat{A}$ and can even be complex. Since then, weak value has been and is constantly being studied extensively both theoretical and experimental points of view. Particularly, the cases of nonlocal observables are interesting \cite{KJResch, KJResch1,Lundeen}, including  EPR paradox \cite{Aha,Aha1}, Hardy's paradox \cite{Hardy,Y.AH,Lundeen1,Yokota}, and Cheshire Cat experiment \cite{YAHR,Tobias}.

In relation to these issues, Vaidman has also claimed that the ^^ ^^ product rule" of commuting observables (for the two-state vector formalism) is not true, i.e., $\langle \hat{A}\hat{B}\rangle_{\rm w} = \langle\hat{A}\rangle_{\rm w}\langle\hat{B}\rangle_{\rm w}$ is not always true even for commuting observables \cite{Vaid,Vaid1}. Here, $\langle \ \cdot \ \rangle_{\rm w}$ is the weak value of the observable inside the bra-ket.

The most studies on weak values focus on continuous variable Gaussian distribution of measuring device. Y. Kedem and L. Vaidman, however, recently considered the interaction between a system and a meter qubit \cite{Y.Kedem}, 
where the system (not necessary be a qubit but could be in a higher dimensional Hilbert space) is conditioned by an initial- and a final-state vectors $| \psi \rangle$ and $|\phi\rangle$ \cite{Y.Aharonov}, and the state of the meter qubit is initially prepared to be $ \gamma | 0\rangle_{\rm m} + \bar\gamma |1\rangle_{\rm m} $ ($\gamma$ and $\bar\gamma$ are real numbers satisfying $\gamma^2 + \bar\gamma^2 = 1$), with $\bar\gamma \ll 1$. 

The interaction Hamiltonian is written as 
\begin{align}{\label{Hamiltonian}}
\hat{H} = g(t) \hat A \hat{\mathbf{\Pi}}_{\rm m} \;, \hspace{0.5cm} \int_{t_0}^t g(t){\rm d} t =  g\delta (t-t_0)\;.
\end{align}
where $ \hat{\mathbf{\Pi}}_{\rm m} \equiv |1\rangle_{\rm m} \langle 1|$ denotes the projection operator onto state $|1\rangle_{\rm m} $ , $\hat A$ represents the Hermitian operator corresponding to the observable of the quantum system, and the coupling $g (t)$ is a nonzero normalized function in a finite-time interaction $(t-t_0)$. Here, the coupling constant $g$ can be arbitrarily large.

The final state of the meter qubit after post-selection of $|1\rangle_{\rm m} $ is calculated as 
\begin{align}{\label{final_state}}
\notag &\langle\phi|e^{-ig\hat A |1\rangle_{\rm m} \langle 1|}|\psi\rangle (\gamma|0\rangle_{\rm m} +\bar\gamma|1\rangle_{\rm m} )\\
\notag &= \ \langle\phi|
  \begin{pmatrix}
  1& 0 \\
  0& e^{-ig\hat A}
 \end{pmatrix}
| \psi\rangle (\gamma|0\rangle_{\rm m} +\bar\gamma|1\rangle_{\rm m} )\\
\notag &=
  \begin{pmatrix}
  \langle\phi|\psi\rangle& 0 \\
   0& \langle\phi|e^{-ig\hat A}|\psi\rangle
 \end{pmatrix}
(\gamma|0\rangle_{\rm m} +\bar\gamma|1\rangle_{\rm m} )\\
&= \ \langle\phi | \psi\rangle \biggl[\gamma | 0 \rangle_{\rm m}  + \bar\gamma\ \dfrac{\langle \phi | e^{-ig\hat A}|\psi\rangle}{\langle\phi | \psi\rangle}|1\rangle_{\rm m} \biggr] \;,
\end{align}	
where we have used the basis $|0\rangle_{\rm m}  = {1 \choose 0}$,  $|1\rangle_{\rm m}  = {0 \choose 1}$, and the complex value sandwiched by $\overline{\gamma}$ and $|1\rangle_{\rm m} $ was named ^^ ^^ modular value" \cite{Y.Kedem} of operator $\hat{A}$, which is written as $(\hat{A})_{\rm mod}$, so that 
\begin{align}{\label{modular}}
(\hat{A})_{\rm mod} \equiv \dfrac{\langle\phi | e^{-ig \hat A}|\psi\rangle}{\langle\phi| \psi\rangle}	\;.
\end{align}

The modular value has the same amplification factor $\frac{1}{\langle\phi| \psi\rangle}$ as weak value. Moreover, in some cases, the modular value has a close association to the weak value. Let us give an example of a spin operators $\hat{\sigma}_x, \hat{\sigma}_y$ and $\hat{\sigma}_z$ and coupling constant $g = -\frac{\pi}{2}$.   We have \cite{Y.Kedem}:
\begin{align}{\label{modular_si}}
(\hat{\sigma})_{\rm mod} \equiv \dfrac{\langle\phi | e^{i\frac{\pi}{2}\hat{\sigma}}|\psi\rangle}{\langle\phi| \psi\rangle}=i\langle \hat{\sigma}\rangle_{\rm w} \quad (\hat{\sigma} = \hat{\sigma}_x, \hat{\sigma}_y \ {\rm or} \ \hat{\sigma}_z)\;.
\end{align}
So, the modular value of a spin component is directly related to its weak value in this specific case.  Interestingly, the modular values can be obtained even for strong coupling constant.

One can raise a significant question: Can we implement this modular-value measurement nondestructive while keeping the system-meter interaction strong? The answer is yes, and such implementation is realized by preparing the meter qubit (or an ensemble of meter qubits) with small $\bar\gamma \ (\ll 1)$ while keeping the coupling constant medium. This choice apparently constraints the expectation value of the interaction Hamiltonian very small, which means that this measurement does not disturb the system's free evolution. Here, one can image that among a huge number of meter-qubit measurements (i.e., events of $|0\rangle_{\rm m}$'s and $|1\rangle_{\rm m}$'s), there are a very little portion of $|1\rangle_{\rm m}$ events, which lead to the actual interaction with the quantum system.

The fundamentals of modular values, however, are not fully understood yet. For example, a general expression for the transformation between modular values and weak values is still missing.  (Note that modular values can be directly related to the weak valuers only for some specific cases such as $\hat A = \hat \sigma$ and $\hat \Pi$ with specific coupling constant $ g = \frac{\pi}{2}$ or $\pi $ in \cite{Y.Kedem}). It might also be an interesting question how to implement a modular-value measurement in an ancillary qubit manner. 

In this paper, we extend the expression Eq. \eqref{modular_si} to the one that relates a modular value to a weak value for more general case, i.e., for an arbitrary observable and coupling constant. Using the extended formalism, the ^^ ^^ failure of sum rule" for modular values is firstly demonstrated.  We also display some examples in which the failure of sum rule for modular values is closely related to the ^^ ^^ failure of product rule" \cite{Vaid,Vaid1} for weak values. Our general method allows us to implement both local and nonlocal measurement, and explains the anomalous results of some interesting experiments. Additionally, a simple quantum circuit that simulates the modular value has been examined, where a controlled rotation gate plays a role of modular value interaction $\hat{U}$ and the half rotation angle stands for the coupling constant $g$.

This paper is organized as follows. The general expression to relate the weak value and modular value is shown in Sec. \ref{s2}. In Sec. \ref{s3}, we introduce the proof of the failure of sum rule, and give some examples of the failure of sum rule in some gedanken or real experiments such as EPR paradox, Hardy paradox, and Cheshire Cat experiments. In Sec. \ref{s4} we consider a controlled-$R_z(\theta)$ gate, where the system qubit controls the meter qubit, to realize the measurement of the modular value of $\sigma_z$ of the system qubit. The paper concludes with remarks in Sec. \ref{s5}.

\section{GENERAL EXPRESSION FOR MODULAR VALUES}{\label{s2}
Our first main result is that, when the dimension of the Hilbert space is two, the weak value for an arbitrary observable can be calculated from its modular value, and vice versa. Let us first start from the case where the dimension ($\equiv n$) of the system Hilbert space is arbitrary but finite (i.e., $n$ can be $\ge 2$). We also assume that the observable $\hat{A}$ has $n$ different eigenvalues $\lambda_k \ (k= 1, 2,...,n)$, which are known. We now introduce the Lagrange interpolation of the matrix form \cite{Li} 
\begin{align}{\label{Lagrange}}
e^{-ig A}= \sum_{k=1}^n e^{-ig\lambda_k}\prod_{\ell=1,\ell\ne k}^n \dfrac{A-\lambda_{\ell} I}{\lambda_k-\lambda_{\ell}}\:,
\end{align}
where $A$ is the matrix expression of $\hat{A}$, and $I$ is the unit matrix. Taking the eigenvectors of $\hat{A}$ as the bases for the matrix expression, Eq.(\ref{Lagrange}) immediately leads to the interpolation of operator form as
\begin{align}{\label{Lagrange2}}
e^{-ig\hat A}= \sum_{k=1}^n e^{-ig\lambda_k}\prod_{\ell=1,\ell\ne k}^n \dfrac{\hat A-\lambda_{\ell}\hat I}{\lambda_k-\lambda_{\ell}}\:. 
\end{align}

Particularly for $n = 2$, this explicitly yields
\begin{align}{\label{Lagrange_3}}
\notag	e^{-ig\hat A}&= e^{-ig\lambda_1} \dfrac{\hat A-\lambda_2\hat I}{\lambda_1-\lambda_2}+e^{-ig\lambda_2} \dfrac{\hat A-\lambda_1\hat I}{\lambda_2-\lambda_1}\\
\notag	&= \dfrac{e^{-ig\lambda_1}-e^{-ig\lambda_2}}{\lambda_1-\lambda_2}\hat A -  \dfrac{\lambda_2e^{-ig\lambda_1}-\lambda_1e^{-ig\lambda_2}}{\lambda_1-\lambda_2}\hat I \\
&= a\hat A + b \hat I \;,
\end{align}
where $a \equiv \dfrac{e^{-ig\lambda_1}-e^{-ig\lambda_2}}{\lambda_1-\lambda_2}$ and $ b \equiv 	-  \dfrac{\lambda_2e^{-ig\lambda_1}-\lambda_1e^{-ig\lambda_2}}{\lambda_1-\lambda_2} $ are complex numbers. Applying pre- and post-selected states, $|\psi\rangle$ and $\langle\phi|$, from right and left, respectively, we obtain the modular value of $\hat{A}$ in relation to its weak value as  
\begin{align}{\label{modular_w}}
(\hat A)_{\rm mod} = a \langle \hat A\rangle_{\rm w} + b \;.
\end{align}
Inversely solving this, it is straightforward to express the weak value of $\hat{A}$ by its modular value as
\begin{align}{\label{weak_modular}}
\langle \hat A\rangle_{\rm w} =\dfrac{1}{a} \biggl ((\hat A)_{\rm mod} - b\biggr) \;.
\end{align}

As the first illustration, let us check whether this reproduces the relation between the weak value and the modular value of a spin operator $\hat{\sigma} \ (= \hat{\sigma}_x, \hat{\sigma}_y$, or $\hat{\sigma}_z)$ in the case of $g = -\frac{\pi}{2}$. The spin operator has two eigenvalues: $\lambda_1 = 1$ for $|\uparrow\rangle$ and $\lambda_2 = -1$ for $|\downarrow\rangle$. Then, the modular value of $\hat{\sigma}$ is immediately given by Eq. \eqref{modular_w} as
\begin{align}{\label{modular_sigma}}
\notag(\hat{\sigma})_{\rm mod} &= \dfrac{e^{i\frac{\pi}{2}}-e^{-i\frac{\pi}{2}}}{2}\langle \hat{\sigma} \rangle_{\rm w} + \dfrac{e^{i\frac{\pi}{2}}+e^{-i\frac{\pi}{2}}}{2}\\
&=i\langle \hat{\sigma}\rangle_{\rm w} \;.
\end{align}	
This is exactly the result obtained by  \cite{Y.Kedem} shown as Eq. \eqref{modular_si} of the present paper.

Another example is the projection operator $\hat{\Pi} = |1\rangle\langle 1|$, which has two eigenvalues: $\lambda_1 = 1$ for $|1\rangle$ and $\lambda_2 = 0$ for $|0\rangle$. For $g=-\frac{\pi}{2}$, we have:
\begin{align}{\label{modular_proj}}
\notag(\Pi)_{\rm mod} &= \dfrac{e^{i\frac{\pi}{2}}-1}{1}\langle \Pi\rangle_{\rm w} + 1\\
&=1 - (1-i)\langle \Pi\rangle_{\rm w} \;.
\end{align}

\noindent If we choose $g = \pi$, it also yields the same result as \cite{Y.Kedem}. 

Now, let us discuss the modular value of the sum of a set of observables $\hat A^{(j)}$, where the  
superscript $(j)$ refers  to the $j^{\rm th}$ subsystem ($j = 1, 2, \cdots, N$). For example, in Sec. \ref{s3}, $j=1$ (or 2) means particle 1 (or 2) in the Bohm-EPR example, and $j=1$ (or 2) means polarization (or ^^ ^^ left path or right path?") of a photon for quantum Cheshire cat case. Later, we will relate this to the weak value of the product of observables $\hat A^{(j)}$. We can assume all $\hat A^{(j)}$ commute with each other since they act in the different spaces $\mathcal{H}^{(j)}$ {of the total Hilbert space $\mathcal{H} = \otimes_{j}\mathcal{H}^{(j)}$}. Therefore, we have $e^{\hat{A}+\hat{B}} = e^{\hat{A}}e^{\hat{B}}$ for $N$ variables. (More precisely, we should write this as $e^{\hat{A}\otimes\hat{I}_{\rm B} + \hat{I}_{\rm A}\otimes\hat{B}} = e^{\hat{A}\otimes\hat{I}_{\rm B}} e^{\hat{I}_{\rm A}\otimes\hat{B}}$, but we avoid this complexity unless things become confusing.) \ Thus we obtain, with the help of Eq. \eqref{Lagrange2}, 
\begin{align}{\label{exp_of_sum}}
\notag e^{-ig\sum_{j=1}^N\hat A^{(j)}}&=\prod_{j-1}^N e^{-ig\hat A^{(j)}}\\
&=\prod_{j=1}^N\biggl[\sum_{k=1}^n e^{-ig\lambda_k^{(j)}}\prod_{l=1,l\ne k}^n \dfrac{\hat A^{(j)}-\lambda_l^{(j)}\hat I}{\lambda_k^{(j)}-\lambda_l^{(j)}}	\biggr] \;,
\end{align}
where, we assumed that the dimension $n$ is the same for all $N$ subsystems.Then, considering the case that the rank of each observable is $2$ (i.e., $n=2$ for each $j$), the modular value of the sum is obtained as  
	\begin{align}{\label{modular_sum}}
	\biggl(\sum_j^N \hat A^{(j)}\biggr)_{\rm mod}= \dfrac{\langle\phi|\prod_j^N\biggl(a^{(j)}\hat A^{(j)}+b^{(j)}\hat I^{(j)}\biggr)|\psi\rangle}{\langle\phi|\psi\rangle}\;.
	\end{align}
where Eq. \eqref{Lagrange_3} is used, and $a^{(j)} \equiv \dfrac{e^{-ig\lambda_1^{(j)}}-e^{-ig\lambda_2^{(j)}}}{\lambda_1^{(j)}-\lambda_2^{(j)}}$ and $ b^{(j)} \equiv -\dfrac{\lambda_2^{(j)}e^{-ig\lambda_1^{(j)}}-\lambda_1^{(j)}e^{-ig\lambda_2^{(j)}}}{\lambda_1^{(j)}-\lambda_2^{(j)}} $.

\section{THE FAILURE OF SUM RULE}{\label{s3}
In quantum mechanics, expectation value of the sum of nonlocal variables is equal to the sum of individual expectation values. This rule also holds for weak values, i.e., $  \langle \sum_j A^{(j)} \rangle_{\rm w} =  \sum_j \langle A^{(j)}\rangle_{\rm w}$. It is worth to note that the sum rule does not hold for modular values. We call this ``the failure of sum rule," which is stated as follows.

\textit{In general, the modular value of sum of observables $\hat A^{(j)}$ is not equal to the sum of the modular values of various $j$}
\begin{align}{\label{modular_sum_modular}}
\biggl(\sum_j^N\hat A^{(j)}\biggr)_{\rm mod} \neq \sum_j^N(\hat A^{(j)})_{\rm mod} \;.
\end{align}	

\textit{Proof}:
From mathematical point of view, we see that $\exp(-ig\sum_j \hat A^{(j)} )\neq \sum_j \exp(-ig\hat A^{(j)})$ for any set of variables $\hat{A}^{(j)}$. Applying pre- and post-selected states from the both sides of the exponential expression, we obtain Eq. \eqref{modular_sum_modular}.

It is also worth to note that the (failure of the) product rule for weak values has some connection with the (failure of the) sum rule for modular values. In fact, considering the case where the product rule works for specific weak values, i.e. $\langle \hat{A}\hat{B}\rangle_{\rm w} = \langle \hat{A}\rangle_{\rm w} \langle \hat{B}\rangle_{\rm w}$, then we can derive that the sum rule also works for their modular values (with additional assumptions), as follows. From Eq. \eqref{modular_sum}, we have
\begin{align}{\label{sum_AB}}
\notag	(\hat A+\hat B)_{\rm mod} &= \dfrac{\langle\phi|\biggl( (a\hat A +b\hat I)(a'\hat B +b'\hat I)\biggr)|\psi\rangle}{\langle\phi|\psi\rangle}\\
\notag	&= aa' \langle \hat A\hat B\rangle_{\rm w}  + ab' \langle \hat A\rangle_{\rm w} + a'b\langle \hat B\rangle_{\rm w} + bb' \\
\notag	&= aa' \langle \hat A\rangle_{\rm w}\langle \hat B\rangle_{\rm w}  + ab' \langle\hat A\rangle_{\rm w} + a'b\langle\hat B\rangle_{\rm w} + bb' \\
\notag	&= (a\langle \hat A\rangle_{\rm w}+b)(a'\langle \hat B\rangle_{\rm w}+b')\\
		&=(\hat A)_{\rm mod}(\hat B)_{\rm mod} \;,
\end{align}
where, $\langle \hat{A}\hat{B}\rangle_{\rm w} = \langle \hat{A}\rangle_{\rm w} \langle \hat{B}\rangle_{\rm w}$ was used. 
Now, if we consider an additional condition, $(\hat A)_{\rm mod}(\hat B)_{\rm mod} = (\hat A)_{\rm mod}+(\hat B)_{\rm mod}$ , which is actually possible \cite{list},  then we obtain $(\hat A+\hat B)_{\rm mod} = (\hat A)_{\rm mod}+(\hat B)_{\rm mod}$. Nevertheless, as was discussed by Vaidman, $\langle \hat{A}\hat{B}\rangle_{\rm w} = \langle \hat{A}\rangle_{\rm w} \langle \hat{B}\rangle_{\rm w}$ is rare to happen in physical world, and so is $(\hat A+\hat B)_{\rm mod} = (\hat A)_{\rm mod}+(\hat B)_{\rm mod}$ as well. 

Hereafter, we provide illustrative examples one by one, where the failure of product rule holds and therefore, the failure of sum rule holds as well.

Let us consider the EPR-Bohm argument \cite{Aha,Aha1}.  Assume that Alice and Bob initially share a maximally entangled singlet state
\begin{align}{\label{singlet_state}}
|\psi\rangle = \dfrac{1}{\sqrt{2}}(|\uparrow_z\rangle_1 |\downarrow_z\rangle_2 - |\downarrow_z\rangle_1 |\uparrow_z\rangle_2) \;,
\end{align}
and postselect the state to $|\phi\rangle = |\uparrow_y\rangle_1 |\uparrow_x\rangle_2${ , where, subscript $j (=1, 2$) means particle $j$.} Then the failure of product rule for weak values is easily seen as follows:
 \begin{align}{\label{sigma_fail_prod}}
\langle\hat{\sigma}_x^{(1)}\rangle_{\rm w}=-1, \hspace{0.2cm} \langle\hat{\sigma}_y^{(2)}\rangle_{\rm w}=-1, \hspace{0.2cm} {\rm but} \hspace{0.2cm} \langle\hat{\sigma}_x^{(1)}\hat{\sigma}_y^{(2)}\rangle_{\rm w}=-1{,}
\end{align}
where superscript $(j)$ also means particle $j$. This result directly leads to the failure of sum rule for the modular values for an arbitrary value of $g$ as
\begin{align}{\label{sigma_fail_sum_1}}
(\hat{\sigma}_x^{(1)})_{\rm mod} = \cos(g) - i\sin(g)\langle\hat{\sigma}_x^{(1)}\rangle_{\rm w} &= \cos(g) + i\sin(g) \;, \\ {\label{sigma_fail_sum_2}}  
(\hat{\sigma}_y^{(2)})_{\rm mod} = \cos(g) - i\sin(g)\langle\hat{\sigma}_y^{(2)}\rangle_{\rm w} &= \cos(g) + i\sin(g) \;, \\ {\label{sigma_fail_sum_3}}
(\hat{\sigma}_x^{(1)}+\hat{\sigma}_y^{(2)})_{\rm mod} &= 1+i\sin(2g) \;.
\end{align}
Clearly, the sum of Eqs.\eqref{sigma_fail_sum_1} and \eqref{sigma_fail_sum_2} is not equal to \eqref{sigma_fail_sum_3} for any value of $g$, which exemplifies the failure of sum rule. Here, Eqs.\eqref{sigma_fail_sum_1} and \eqref{sigma_fail_sum_2} are derived from Eq.\eqref{modular_w} by putting $\lambda_1 = 1$ and $\lambda_2 = -1$, and Eq.\eqref{sigma_fail_sum_3} is obtained as follows. We start from
\begin{widetext}
\begin{align}{\label{calc_sigma_fail_sum_3}}
\notag e^{-ig \left(\hat{\sigma}_x^{(1)} + \hat{\sigma}_y^{(2)}\right)} &= e^{-ig \hat{\sigma}_x^{(1)}} \otimes e^{-ig \hat{\sigma}_y^{(2)}} 
= \left(\cos(g) \ \hat{I}^{(1)} - i\sin(g) \ \hat{\sigma}_x^{(1)}\right)\otimes  \left(\cos(g) \ \hat{I}^{(2)} - i\sin(g) \ \hat{\sigma}_y^{(2)}\right) \\ 
& = \cos^2(g) \ \hat{I}^{(1)} \otimes \hat{I}^{(2)} 
- i \cos(g) \sin(g) \left( \hat{I}^{(1)} \otimes \hat{\sigma}_y^{(2)} + \hat{\sigma}_x^{(1)} \otimes \hat{I}^{(2)}  \right) 
- \sin^2(g) \ \hat{\sigma}_x^{(1)} \otimes \hat{\sigma}_y^{(2)}  \;, 
\end{align}
\end{widetext}
and applying $\langle \phi|$ from the left and $|\psi\rangle$ from the right, we obtain
\begin{widetext}
\begin{align}{\label{calc_sigma_fail_sum_4}}
(\hat{\sigma}_x^{(1)}+\hat{\sigma}_y^{(2)})_{\rm mod} 
&=  \cos^2 (g) -  {i \over 2}\sin(2g) \left( \langle \hat{\sigma}_y^{(2)} \rangle_{\rm w} + \langle \hat{\sigma}_x^{(1)}\rangle_{\rm w}  \right) 
- \ \sin^2 (g) \ \langle \hat{\sigma}_x^{(1)}\rangle_{\rm w} \langle \hat{\sigma}_y^{(2)} \rangle_{\rm w}  \;, 
\end{align}
\end{widetext}	
which immediately yields Eq.\eqref{sigma_fail_sum_3}. In this way we can directly obtain the failure of sum rule for the modular values. Actually, we can follow the calculation in reverse. This means that the failure of product rule for the weak values can be directly derived from the failure of sum rule for the modular values for arbitrary values of $g$.

The second example is Hardy's paradox experiments \cite{Hardy,Y.AH,Lundeen1,Yokota}. In this setup (Fig.\ref{fig_Hardy}), a monochromatic electron and a positron are respectively put into each Mach-Zehnder interferometer. This initial condition leads to the state 
\begin{align}{\label{psi_0}}
\notag|\psi_0\rangle = \dfrac{1}{\sqrt{4}}\biggl(&|O\rangle_+|NO\rangle_-+|NO\rangle_+|O\rangle_-\\
&+|NO\rangle_+|NO\rangle_- +|O\rangle_+|O\rangle_-\biggr) \;,
\end{align} 
just after passing through BS1. Here state $|O\rangle_-$ $\left( |O\rangle_+ \right)$ denotes the electron (positron) that goes to the overlapping region, and the states $|NO\rangle_-$ $\left( |NO\rangle_+ \right)$ denotes the electron (positron) that goes to the non-overlapping region. The subsystem subscript $j$ thus represents $j=+$ (positron) or $-$ (electron) instead of $j=1$ or 2. After the pairwise-extinction point, the possibility of $|O\rangle_+|O\rangle_-$ vanishes, which leads to the prepared state 
\begin{align}{\label{Hardy_state2}}
|\psi\rangle &= \dfrac{1}{\sqrt{3}}\biggl(|O\rangle_+|NO\rangle_-+|NO\rangle_+|O\rangle_-+|NO\rangle_+|NO\rangle_-\biggr) \;,
\end{align}
where the denominator $\sqrt{4}$ is renormalized into $\sqrt{3}$. This is the pre-selected state. 

\begin{figure}
\centering
\includegraphics[width=8.6 cm]{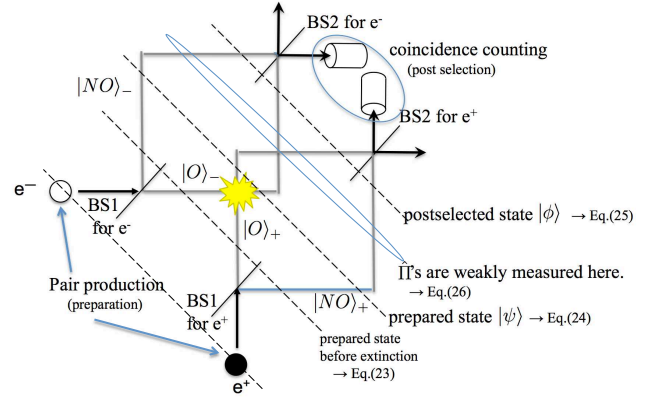}
\caption{(Color online) The set-up for Hardy's paradox.
}
\label{fig_Hardy}
\end{figure}

Now, we postselect the case that two detectors in Fig.\ref{fig_Hardy} click simultaneously. Tracing back to the point before BS2s, this post-projection is attributed to the expression of the postselected state $|\phi\rangle$ as
\begin{align}{\label{Hardy_state2}}
|\phi\rangle &= \dfrac{1}{2}\biggl(|O\rangle_+ - |NO\rangle_+\biggr)\biggl(|O\rangle_- - |NO\rangle_-\biggr)  \;.
\end{align}
So, by preparing an electron and a positron at the same time and selecting the case that two detectors have coincident counting, we can realize the projection of $|\psi\rangle$ onto $|\phi\rangle$ inside the interferometer. 
The ^^ ^^ which path?" measurement can be weakly measured between  $|\psi\rangle$ and $|\phi\rangle$, which gives the weak values of the projection operators.  

The product rule does not hold in Hardy's case neither as 
\begin{align}{\label{Hardy_fail_prod}}
\langle\hat{\Pi}_{O^+}\rangle_{\rm w}=1, \hspace{0.2cm} \langle\hat{\Pi}_{O^-}\rangle_{\rm w}=1, \hspace{0.2cm} {\rm but} \hspace{0.2cm} \langle\hat{\Pi}_{O^+}\hat{\Pi}_{O^-}\rangle_{\rm w}=0 \;,
\end{align}
which was predicted by Aharonov et.al.,\cite{Y.AH} and experimentally verified by Lundeen et.al.,\cite{Lundeen1} and Yokota et.al., \cite{Yokota}. This indeed is the observation that weak values well explain the paradoxical behavior of the individual probabilities and their joint probability.  Now, by performing the same procedure as the first example (EPR paradox), it is easy to verify the failure of sum rule as follow:
	\begin{align}{\label{sigma_fail_sum}}
	&(\hat{\Pi}_{O^+})_{\rm mod} = (e^{-ig}-1)\langle\hat{\Pi}_{O^+}\rangle_{\rm w}+1 = e^{-ig} \;, \\
	&(\hat{\Pi}_{O^-})_{\rm mod} = (e^{-ig}-1)\langle\hat{\Pi}_{O^-}\rangle_{\rm w}+1 = e^{-ig} \;, \quad {\rm and}\\
	&(\hat{\Pi}_{O^+}+\hat{\Pi}_{O^-})_{\rm mod} = 2e^{-ig}-1\;.
	\end{align}	
Obviously, the modular value of sum $(\hat{\Pi}_{O^+}+\hat{\Pi}_{O^-})_{\rm mod}$ is different from the sum of modular values $(\hat{\Pi}_{O^+})_{\rm mod} + (\hat{\Pi}_{O^-})_{\rm mod}$ by $-1$ (for any $g$).

Hereafter, let us apply our method to the analysis of the quantum Cheshire cat as the third example. The concept of quantum Cheshire cat is given in \cite{YAHR}, and experimentally verified in \cite{Tobias}. In this experiment, a quantum particle (neutrons are used in \cite{Tobias} but any qubits are OK such as photons) having spin (or polarization if photons are used) is compared to the cat with two possibilities of its paths $|L\rangle$ and $|R\rangle$ of a Mach-Zehnder interferometer and its spins $|\uparrow\rangle$ and $|\downarrow\rangle$. The ^^ ^^ $|L\rangle$ or $|R\rangle$?" information corresponds to the position of the cat body, and spin is considered to be the cat's grin.
By putting a particle-beam attenuator in path $L$ or $R$, or applying a magnetic field which changes the particle's spin, 
Hasegawa's group succeeded in observing that the particle (the body of the cat) goes through one path whereas its spin (the cat's grin) goes through the other path \cite{Tobias}. 

In this work, we follow the notation of \cite{YAHR}, i.e., we assume the initial state of}
the single photon to be
	\begin{align}{\label{psi_Cheshire}}
	|\psi\rangle = \dfrac{1}{\sqrt{2}}|H\rangle_1 \biggl( i |L\rangle_2 + |R\rangle_2\biggr) \;,
	\end{align}	
which can be prepared by sending a horizontally polarized photon toward a 50/50 beam splitter (BS1 in Fig.\ref{fig_Cheshire}). Here, subscript 1 $(j=1)$ denotes the polarization degree of freedom, and 2 $(j=2)$ denotes that of the ^^ ^^ which path?". The use of these suffixes that look seemingly redundant is a stepping stone for the later introduction of the meter qubit. The quantum Cheshire cat is observed when we perform weak measurement on ^^ ^^ which-path?" in one arm and weak measurement on the polarization in the other arm conditioned by the subsequent post-selection of the projection onto state $|\phi\rangle$:
	\begin{align}{\label{phi_Cheshire}}
	|\phi\rangle \equiv 
	\dfrac{-i}{\sqrt{2}} \biggl(|H\rangle_1 |L\rangle_2 +  |V\rangle_1 |R\rangle_2 \biggr) \;.
	\end{align}

The result of the weak measurement, as is described in the below, suggests that the particle travels along one path, whereas its polarization goes along the other. This strange result can be well expressed by the weak values of the measurement. The calculation of the weak values for projection operators $\hat \Pi_j ^{(2)}= |j\rangle\langle j|$\ ($j = L, R $) yields
\begin{align}{\label{weak_LR}}
\langle \hat{\Pi}_L^{(2)}\rangle_{\rm w} = 1\hspace{0.5cm} {\rm and} \hspace{1cm}\langle \hat{\Pi}_R^{(2)}\rangle_{\rm w} = 0 \;.
\end{align}

\begin{figure}
\centering
\includegraphics[width=8.6 cm]{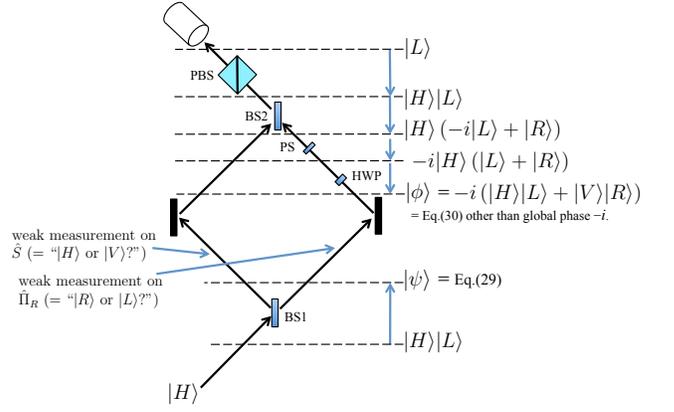}
\caption{(Color online) Set-up for Cheshire Cat.
}
\label{fig_Cheshire}
\end{figure}

This means that the particle just travels on the left side. Note that these local weak measurements can be performed simultaneously and the result at one location does not affect the result at the other location. Next we carry out {the calculation for} the nonlocal weak measurement to determine the location of polarization component which yields
\begin{align}{\label{weak_sigma_LR}}
\langle \hat{S}^{(1)}\hat{\Pi}_L^{(2)}\rangle_{\rm w} = 0\hspace{0.5cm} {{\rm and} \hspace{1cm}} \langle \hat{S}^{(1)}\hat{\Pi}_R^{(2)}\rangle_{\rm w} = 1 \;.
\end{align}
where $\hat S = |H\rangle\langle H| -  |V\rangle\langle V|$ is one of the Stokes operators for polarization. This implies that the polarization component of quantum particle located in the right side. The Cheshire cat really exist in quantum world!

Now, we show that the above results can be directly obtained from measurement of the modular values. We consider a nonlocal meter consisting entangled two qubits initially in the state:
\begin{align}{\label{2qubit state}}
\gamma |HH\rangle + \bar\gamma [|HV\rangle + |VH\rangle + |VV\rangle] \;,
\end{align}
where $\gamma^2 + 3\bar\gamma^2 = 1$ and $\bar\gamma \ll  1$. More in detail, we put suffixes 1m and 2m for the meter photons as 
\begin{align}{\label{2meter qubit state}}
\notag &\gamma |H\rangle_{\rm 1m}|H\rangle_{\rm 2m} + \bar\gamma [|H\rangle_{\rm 1m}|V\rangle_{\rm 2m} \\
&+ |V\rangle_{\rm 1m}|H\rangle_{\rm 2m} + |V\rangle_{\rm 1m}|V\rangle_{\rm 2m}] \;,
\end{align}
We assume that the polarization of the system photon is measured by the polarization of the meter photon 1m, and ^^ ^^ which path?" information of the system photon is measured by the polarization of the meter photon 2m. This can be done by the Hamiltonian 
	\begin{align}{\label{Hamil}}
	\hat H = g
	\biggl(\hat S^{(1)}\hat{\Pi}_V^{(1{\rm m})} +\hat\Pi_{L(R)}^{(2)}\hat \Pi_V^{(2{\rm m})} \biggr) \;,
	\end{align}
where $\hat\Pi_V^{(j{\rm m})} = |V\rangle\langle V|$ is projection operator. Using Eqs. (\ref{modular_sigma}, \ref{modular_proj}) and Eq. \eqref{modular_sum} we have :
\begin{align}{\label{modular_sm/LR}}
(\hat{\Pi}_{L(R)}^{(2)})_{\rm mod} &= (e^{-ig}-1)\langle \hat{\Pi}_{L(R)}^{(2)}\rangle_{\rm w} +1\\
(\hat{S}^{(1)})_{\rm mod} &= e^{-ig} \langle \hat{S}^{(1)}\rangle_{\rm w}\\
\notag(\hat{S}^{(1)}+\hat{\Pi}_{L(R)}^{(2)})_{\rm mod}&=e^{-ig}\Bigl[\langle \hat{S}^{(1)}\rangle_{\rm w} \\
&-(e^{-ig}-1)\langle \hat{S}^{(1)}\hat{\Pi}_{L(R)}^{(2)}\rangle_{\rm w} \Bigr]\;.
\end{align}
In other words, we can directly derive weak values by measuring the modular values.

In the experiment, one can perform the modular-value measurement for pre- and post-selected ensemble using the meter qubit prepared in the state \eqref{2qubit state}. The final state of the qubit becomes
\begin{widetext}
\begin{align}{\label{final_2}}
\langle\phi|\psi\rangle\biggl\{\gamma|HH\rangle+\bar\gamma\biggl[(\hat{S}^{(1)})_{\rm mod}|HV\rangle+(\hat{\Pi}_{L(R)}^{(2)})_{\rm mod}|VH\rangle+\bigl(\hat{S}^{(1)}+\hat{\Pi}_{L(R)}^{(2)}\bigr)_{\rm mod}|VV\rangle\biggr]\biggr\} \;.
\end{align}
\end{widetext}
Therefore, by performing the tomography on the final state, one can obtain the modular values.

Lastly, it might be reasonable to compare our approach with the concept of ``contextual value" which were introduced by Dressel \textit{et al.} \cite {Dressel, Dressel1}. At first sight, both methods can yield weak values, and generalize the measurement of observables. More specifically, the contextual value in \cite{Dressel, Dressel1} corresponds to AAV effect. It also can yield the direct result \cite{Pryde} for a QND measurement. However, our exploration about the failure of sum rule for modular values could bring us to close to understand the non local characters of quantum mechanics. \\

\section{C-$R_\text{z}(\theta)$ GATE AS MODULAR VALUE MEASUREMENT}{\label{s4}
We consider a simple quantum circuit that implement the modular value. In this scheme, an ancillary system qubit couples to a meter qubit by the controlled Z gate ($=$ C-$R_\text{z}(\theta)$ gate) so that the meter qubit is controlled by the system qubit as shown in the inset of Fig \ref{fig1}. The rotation Z gate $R_z(\theta)$ has the form $R_z(\theta) = e^{-i\theta\hat\sigma_z/2} = \text{diag}(e^{-i\theta/2}, e^{i\theta/2})$ \cite{Nielsen}, where we have used the eigenvalues 1 and $-1$, and eigenfunctions $|\uparrow\rangle$ and $|\downarrow\rangle$ of  Pauli matrix $\hat\sigma_z$, respectively. Therefore, the C-$R_\text{z}(\theta)$ can simulate the unitary $\hat U$ with $g = \theta / 2$, and $\hat A = \hat\sigma_z$ as follow: 
	\begin{align}{\label{usm}}
	\notag \hat U&= e^{-i\theta\hat\sigma_z\otimes\hat P/2}\\
	&= \hat I\otimes |0\rangle\langle 0| + e^{-i\theta\hat\sigma_z/2}\otimes |1\rangle\langle 1|\;.
	\end{align}
where we have used $\hat P = |1\rangle\langle 1|$ (or, in general, $\hat P = 0|0\rangle\langle 0| + 1|1\rangle\langle 1|$), and the last line of Eq. \eqref{usm} shows the  C-$R_z(\theta)$ gate operation.

In the below, we describe the case where the pre- and post-selected states for the system are chosen as $|\psi\rangle = (|\uparrow\rangle + |\downarrow\rangle)/\sqrt{2}$ and $|\phi\rangle = [\sqrt{2+\sqrt{2}} |\uparrow\rangle - \sqrt{2-\sqrt{2}}|\downarrow\rangle]/2$ \cite{Aharonov2005}. It is straightforward to calculate the weak value and modular value of $\hat\sigma_z$ for arbitrary values of $\theta (= 2g)$, resulting in $\langle\sigma_z\rangle_{\rm w} = 1+\sqrt{2}$ and $(\sigma_z)_{\rm mod} = \cos(\theta/2) + i( 1+\sqrt{2})\sin(\theta/2)$, which is shown in Fig. \ref{fig1}.  Obviously, in this example, both weak and modular values lie outside the range of eigenvalues of $\hat\sigma_z$, and particularly, the modular value becomes complex. Nevertheless, the modulus of modular value lies between the expectation value and weak value of observable $\sigma_z$ such as $1 \le |(\sigma_z)_{\rm mod}| \le 1+\sqrt{2}$ as shown in Fig. \ref{fig1}. Additionally, modular value becomes weak value when $g = \pi/2$ (the center peak in Fig. \ref{fig1}).

One might naively think that when the coupling constant is made sufficiently small, the modular value will converge to the weak value. This is, however, incorrect. In fact, using the Taylor series expansion up to the first order of coupling constant, we obtain $(\sigma_z)_{\rm mod} = 1 -ig\langle \sigma_z\rangle_{\rm w}\approx 1$ [see Fig. \ref{fig1}]. Therefore, in considering the modular value with small coupling constant, we need to calibrate to obtain the correct weak value. Hopefully, this example can be tested in the laboratory with the aid of the current quantum information technology, which can easily be implemented by the a controlled rotation gates. 

\begin{figure}
\centering
\includegraphics[width=8.6 cm]{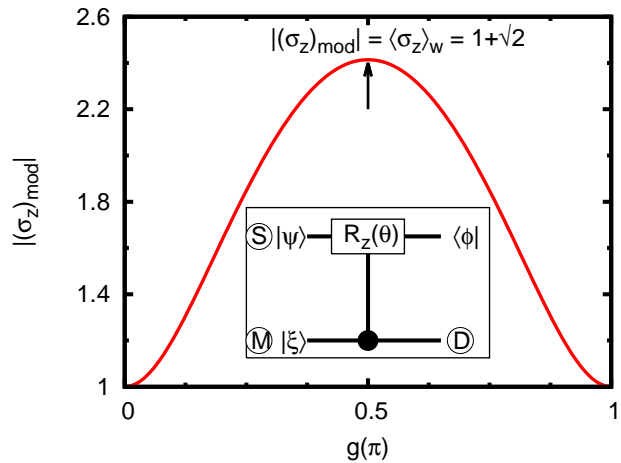}
\caption{(Color online) Main figure: Modulus of modular value as a function of coupling constant $g = \theta / 2$. The highest position, which is shown by the black arrow, corresponds to weak value when $g = \pi/2$. Inset: Quantum circuit simulates the modular value. The system is prepared in state $|\psi\rangle$ and post-selected to be $|\phi\rangle$. The meter qubit state is prepared as $|\xi\rangle = \gamma|0\rangle +\bar\gamma|1\rangle$, and measured in $\hat\sigma_z$ basis. S (or M) presents the System (or qubit Meter) respectively, the outcome is read out by the Detector D.
}
\label{fig1}
\end{figure}

\section{CONCLUSIONS}{\label{s5}
In conclusion, we showed that the modular values and weak values of an observable are closely related via the Lagrange interpolation formula. For two-dimensional cases, particularly, they can be simply derived from each other via Eqs. (\ref{modular_w}, \ref{weak_modular}). It enables one to obtain weak values by experimentally obtaining the modular values, which do not require infinitesimally small coupling. Similarly to the failure of product rule of weak values \cite{Vaid, Vaid1}, we also showed the failure of sum rule of modular values. This also gives a new way of explanation of paradoxes, which we described through a number of examples such as EPR argument, Hardy's paradox, and quantum Cheshire Cat. Lastly, we gave a simple implementation of measuring modular values with a C-$R_\text{z}(\theta)$ gate, which might allow us to measure modular values in a practical way.

\begin{acknowledgments}
This work was supported by JSPS Grant-in-Aid for Scientific Research(A) 25247068.
\end{acknowledgments}

\bibliography{basename of .bib file}

\begin{thebibliography}{99}
\bibitem {list} There are many cases that satisfy this condition. We list here from -10 to 10 with step 0.1 $((A)_{\rm m},(B)_{\rm m})$ or vice versa: ($-1,\frac{1}{2}$); ($-i, \frac{1}{2} + \frac{1}{2} i$); (0,0); ($i, \frac{1}{2} - \frac{1}{2} i$); ($1-2 i, 1 + \frac{1}{2} i$); ($1-i, 1+  i$); ($1-\frac{1}{2}i, 1+ \frac{1}{2} i$); ($\frac{3}{2}, 3$); ($\frac{3}{2}+\frac{1}{2}i, 2- i$); (2,2).
\end{thebibliography}

\end{document}